\documentstyle[11pt]{article}
\begin{document}

\title{A New Kind of Graded Lie Algebra and Parastatistical Supersymmetry
\thanks{This project supported by National Natural Science Foundation of china
and LWTZ 1298.}}
\author{{Yang Weimin and Jing Sicong}\\
{\small Department of Modern Physics, University of Science and }\\
{\small Technology of china, Hefei 230026}}
\maketitle
\bigskip
\begin{abstract}
In this paper the usual $Z_2$ graded Lie algebra is generalized to a new form,
which may be called $Z_{2,2}$ graded Lie algebra. It is shown that there
exists close connections between the $Z_{2,2}$ graded Lie algebra and
parastatistics, so the $Z_{2,2}$ can be used to study and analyse various
symmetries and supersymmetries of the paraparticle systems.
\end{abstract}
\bigskip

\section{Introduction}
\par
It is well known that the symmetry plays a fundamental role in theoretical
physics research. When a new symmetry of investigating system is revealed, it
will lead not only to better understanding of the system but sometimes give
rise to establishing unexpected and important relations between different
theories and lines of research. Parastatistics first was introduced as an
exotic possibility extending the Bose and Fermi statistics \cite{s1}, and for
the long period of time the interest to it was rather academic. Recently it
finds applications in the physics of the quantum Hall effect and it probably
is relevant to high temperature superconductivity \cite{s2}, so it draws more
and more attentions from both theoretical and experimental physicists.
Supersymmetry, established in the early 70's, unifies Bose and Fermi symmetry
and leads to deeply developments of field and string theories which become the
cornerstones of modern theoretical physics. A natural question is, can we
unify parabosons and parafermions into some kind of supersymmetric theories?
Though some recent researches indicated that the so-called parasupersymmetry
(paraSUSY) maybe unifies these paraparticles \cite{s3}, nevertheless, the two
concepts, i.e., the parastatistics and the supersymmetry, seem to be still
independent from the point of view of algebraic structure. In fact, the
mathematical basis of usual supersymmetry is $Z_2$ graded Lie algebra
\cite{s4}, in which not only commutators but also anticommutators are
involved. However, the characteristic of algebraic relations of parastatistics
is trilinear commutation relations, or double commutation relations. Thus two
paraparticles neither commute nor anticommute with each other, and in place of
this, three paraparticles satisfy some complicated commutation relations,
which certainly bring some intrinsic difficulties into the research of
paraSUSY.
\par
A new kind of graded Lie algebra (we call it $Z_{2,2}$ graded Lie algebra) is
introduced in this paper to provide a suitable mathematical basis for the
paraSUSY theories. This $Z_{2,2}$ graded Lie algebra is a natural extension or
generalization of the usual $Z_2$ one. Especially, the trilinear commutation
relations of parastatistics can automatically appear in the structure of
$Z_{2,2}$ graded Lie algebra, therefore, supersymmetric theories based on the
$Z_{2,2}$ graded Lie algebra may unify the concepts of parastatistics and
supersymmetry. It can be shown that the algebraic structure of paraSUSY
systems indeed is the $Z_{2,2}$ graded Lie algebra. So we can classify and
analyse the possible symmetries and supersymmetries of paraparticle systems more
systematically and more effectively from the point of view of $Z_{2,2}$ graded
Lie algebra.
\par
The paper is arranged as follows. In section 2 the mathematical definition of
the $Z_{2,2}$ is introduced with some examples showing how to produce a
$Z_{2,2}$ graded Lie algebra from a usual Lie algebra, especially how to
derive the characteristic trilinear commutation relations of parastatistics.
In section 3 we demonstrate that the algebraic structure of systems consisting of
parabosons and parafermions is nothing but the $Z_{2,2}$ graded Lie algebra.
We analyse the various symmetries and supersymmetries of paraparticle systems
in section 4, and give some summary and discussion in the last section.

\section{$Z_{2,2}$ graded Lie algebra} 
\par
Let vector space ${\bf L}$ over a field ${\bf K}$ be a direct sum
of four subspaces $L_{ij}$ (i,j=0,1), i.e.,
\begin{equation}
{\bf L} = L_{00} \oplus L_{01} \oplus L_{10} \oplus L_{11}.
\end{equation}
For any two elements in ${\bf L}$, we define a composition (or
product) rule, written $\circ$, with the following properties
\par
(i) Closure: For $\forall u,v \in {\bf L}$, we have $u \circ v = w
\in {\bf L}$, i.e.,
\begin{equation}
{\bf L} \circ {\bf L} \rightarrow {\bf L}.
\end{equation}
\par
(ii) Bilinearity: For $\forall u,v,w \in {\bf L}$, $c_1,c_2 \in {\bf K}$, we
have
\begin{eqnarray}
(c_1 u + c_2 v) \circ w = c_1 u \circ w + c_2 v \circ w,  \nonumber\\
w \circ (c_1 u + c_2 v) = c_1 w \circ u + c_2 w \circ v.
\end{eqnarray}
\par
(iii) Grading: For $\forall u \in L_{ij}, v \in L_{mn}, (i,j,m,n = 0,1)$, we
have
$$
u \circ v = w \in L_{(i+m) mod 2, (j+n) mod 2}.
$$
i.e.
\begin{equation}
L_{ij} \circ L_{mn} \rightarrow L_{(i+m) mod 2, (j+n) mod 2},
\end{equation}
where $mod 2$ means cutting $2$ when $i+m =2$. For instance,
$L_{00} \circ L_{00} \rightarrow L_{00}$, $L_{10} \circ L_{11}
\rightarrow L_{01}$, $L_{11} \circ L_{11} \rightarrow L_{00}$, $\cdots$.
\par
(iv) Supersymmetrization: For $\forall u \in L_{ij}, v \in L_{mn}$, we have
\begin{equation}
u \circ v = - (-1)^{g(u) \cdot g(v)} v \circ u,
\end{equation}
here we assign to any $u \in L_{ij}$ a degree $g(u) = (i,j)$ which satisfies
$$
g(u) \cdot g(v) = (i,j) \cdot (m,n) = im + jn,
$$
$$
g(u) + g(v) = (i,j) + (m,n) = (i+m,j+n)_{mod 2},
$$
where $v \in L_{mn}$. Obviously, the $g(u)$ looks like a two-dimensional
vector and the above two expressions are exactly dot product and additive
operations of the two-dimensional vectors.
\par
(v) Generalized Jacobi identities: For $\forall u \in L_{ij}, v \in L_{kl},
w \in L_{mn}, (i,j,k,l,m,n =0,1)$, we have
\begin{equation}
u\circ (v \circ w) (-1)^{g(u) \cdot g(w)} + v \circ (w \circ u) (-1)^{g(v)
\cdot g(u)} + w \circ (u \circ v) (-1)^{g(w) \cdot g(v)} = 0.
\end{equation}
It is easily to know that there are totally 20 different possibilities for
constructing the generalized Jacobi identities from 4 subspaces of ${\bf L}$ 
$(C_{4}^{1} + C_{4}^{1} C_{3}^{2} + C_{4}^{3} = 20).$
\par
{\bf Definition}: {\sl A linear space satisfying the above conditions (i)-(v)
is called the $Z_{2,2}$ graded Lie algebra.}
\par
For instance, one can define the product rule on ${\bf L}$ as
\begin{equation}
u \circ v = uv - (-1)^{g(u) \cdot g(v)} vu,
\end{equation}
for $\forall u, v \in {\bf L}$, where the elements $u$, $v$ can be treated as
operators in some Hilbert space and the expression $uv$ can be understood as
a product of the two operators, it is straightforward to check this product
rule satisfying the above conditions (ii), (iv) and (v). Furthermore, if one
impose the closure and the grading on the product rule (7), the all elements
in ${\bf L}$ will form a $Z_{2,2}$ graded Lie algebra according to commutation
or anticommutation relations, and the generalized Jacobi identities will take
the usual trilinear (or double brackets) form.
\par
Thus we give the full definition of the $Z_{2,2}$ graded Lie algebra. It is
easy to see that the $Z_{2,2}$ graded Lie algebra is a directly generalization
or extension of the usual $Z_2$ graded Lie algebra in the following several
aspects: their product rules are the same, and both of them have the three
basic characters of graded Lie algebra, i.e., grading, supersymmetrization and
generalized Jacobi identities. The only difference between them is, for $Z_2$
case, which is a direct sum of two subspaces, the grading is one-dimensional,
and for $Z_{2,2}$ case, which is a direct sum of four subspaces, the grading
is two-dimensional. Correspondingly, the degree of elements in $Z_2$ is only a
number, and that in $Z_{2,2}$ is a two-dimensional vector. It is worth
mentioning that there are only four possibilities to construct the generalized
Jacobi identities in terms of trilinear or double brackets mathematically,
i.e.,
\begin{eqnarray}
&&[A,[B,C]] + [B,[C,A]] + [C,[A,B]] = 0, \nonumber\\
&&[A,\{B,C\}] + [B,\{C,A\}] + [C,\{A,B\}] = 0, \nonumber\\
&&[A,\{B,C\}] + \{B,[C,A]\} - \{C,[A,B]\} = 0, \nonumber\\
&&[A,[B,C]] + \{B,\{C,A\}\} - \{C,\{A,B\}\} = 0.
\end{eqnarray}
Here we want to point out that only the first three expressions of (8) appear
in the generalized Jacobi identities of the usual $Z_2$ graded Lie algebra
\cite{s4}, however, all the four expressions of (8) will appear in the
generalized Jacobi identities of $Z_{2,2}$. This fact indicates that the
$Z_{2,2}$ is more complete in the algebraic structure, and has higher
symmetries than $Z_2$. Their further connection will become clearer later in
the symmetry analysis.
\par
Now we take the Lie algebra $su(1,1)$ as an example to show how to produce the
$Z_{2,2}$ extension of $su(1,1)$. We take $su(1,1)$ as the $L_{00}$ subspace
which is three-dimensional, however, the dimensions of subspaces $L_{01}$,
$L_{10}$ and $L_{11}$ are restricted by the generalized Jacobi identities.
Simple calculations show that in a nontrivial maximum-dimensional extension,
the above three subspaces can only contain 2, 2 and 1 elements respectively.
Using the notations of elememts in different subspaces showing in the Table 1,
we have
\begin{eqnarray}
&& [\tau_1, \tau_2]=-\tau_3, ~~~[\tau_2, \tau_3]=\tau_1, ~~~[\tau_3, \tau_1]
=\tau_2, \nonumber\\
&& [\tau_i, Q_{\alpha}]=-(t_i)_{\alpha \beta} Q_{\beta}, ~[\tau_i, a_m]=
-(c_i)_{mn} a_n, ~[\tau_i, f]=0, \nonumber\\
&& \{Q_{\alpha}, Q_{\beta}\}=(h_i)_{\alpha \beta} \tau_i, ~\{Q_{\alpha}, a_m\}
=\lambda_1 (\sigma_2)_{\alpha m} f, ~\{Q_{\alpha}, f\}=
\lambda_2 (I)_{\alpha m} a_m,  \nonumber\\
&&\{a_m, a_n\}=(d_i)_{m n} \tau_i,~\{a_m, f\}=
\lambda_3 (I)_{m \alpha} Q_{\alpha},
\end{eqnarray}
where $i=1,2,3$, $m,n=1,2$ and $\alpha,\beta=1,2$. The first line in (9) is
the Lie algebraic relations in $L_{00}$ subspace, and the others on (9) are
the algebraic relations of the $Z_{2,2}$ extension. All the structure
constants such as $t_i, c_i, h_i, d_i$ are determined by the generalized
Jacobi identities as
\begin{eqnarray*}
&&t_1=-\sigma_{1}/2, ~t_2=-\sigma_{2}/2, ~t_3=-i \sigma_{3}/2, ~c_i=t_i,
\nonumber\\
&&h_1=-2i\lambda_1 \lambda_2 \sigma_3, ~h_2=-2\lambda_1 \lambda_2 I, ~
h_3=2\lambda_1\lambda_2 \sigma_1,  \nonumber\\
&&d_1=-2i\lambda_1 \lambda_3 \sigma_3, ~d_2=-2\lambda_1 \lambda_3 I, ~
d_3=2\lambda_1 \lambda_3 \sigma_1,
\end{eqnarray*}
where $\sigma_i$ and $I$ are the Pauli matrices and $(2 \times 2)$ unit
matrix respectively. It needs to be explained that the three undetermined
constants $\lambda_i$ may be absorbed into redifinitions of the elements in
$L_{01}$, $L_{10}$ and $L_{11}$ subspaces.
\par
If we consider a unitary representation of the $Z_{2,2}$ extension of the
$su(1,1)$, and take the constants as $\lambda_1 =-i$, $\lambda_2 =0$ and
$\lambda_3 =2$, then under mappings $-2i\tau_3 \rightarrow M(a)$,
$2(-\tau_1 + i\tau_2) \rightarrow B^{\dagger}$,
$2(\tau_1 + i\tau_2) \rightarrow B$,
$a_1 \rightarrow a^{\dagger}$ and
$a_2 \rightarrow a$, we can derive from (9) the following relations
\begin{eqnarray}
&& [M(a), a]=-a, ~~~[B, a]=0, ~~~[B, a^{\dagger}]=2a, \nonumber\\
&& \{a^{\dagger}, a\}=2M(a), ~~~\{a, a\}=2B,
\end{eqnarray}
together with the adjoint ones. Obviously, these relations are exactly
equivalent to the single mode paraboson algebraic relations
\begin{equation}
[\{a^{\dagger}, a\}, a]=-2a, ~~~[\{a, a\}, a]=0,
~~~[\{a, a\}, a^{\dagger}]=4a.
\end{equation}
Thus we see that the parastatistical algebraic relations can be automatically
derived from the $Z_{2,2}$ algebraic relations.

\section{Algebraic structure of parastatistics}
\par
Let us consider a paraparticle system consisting of M-mode parabosons (whose
creation and annihilation operators are denoted by $a_k^{\dagger}$ and $a_k$
respectively) and N-mode parafermions (denoted by $f_{\alpha}^{\dagger}$ and
$f_{\alpha}$), whose algebraic structure can be expressed in terms of the
following 12 independent relations (the Latin indices take values from 1 to M,
and the Greek ones from 1 to N)
\begin{eqnarray}
&& [a_k, \{a_l^{\dagger}, a_m\}]=2\delta_{kl} a_m,~~~[f_{\alpha},
[f_{\beta}^{\dagger}, f_{\gamma}]]=2\delta_{\alpha \beta} f_{\gamma},
\nonumber\\
&& [a_k, \{a_l, a_m\}]=0,~~~[f_{\alpha}, [f_{\beta}, f_{\gamma}]]=0,
\nonumber\\
&& [a_k, [f_{\alpha}^{\dagger}, f_{\beta}]]=0, ~~~[f_{\alpha},
\{a_k^{\dagger}, a_l\}]=0, \nonumber\\
&& [a_k, \{a_l, f_{\alpha}\}]=0, ~~~\{f_{\alpha}, \{a_k, f_{\beta}\}\}=0,
\nonumber\\
&& [a_k, \{a_l^{\dagger}, f_{\alpha}\}]=2\delta_{kl} f_{\alpha}, ~~~
\{f_{\alpha}, \{a_k^{\dagger}, f_{\beta}\}\}=0, \nonumber\\
&& [a_k, \{a_l, f_{\alpha}^{\dagger}\}]=0, ~~~\{f_{\alpha}, \{a_k,
f_{\beta}^{\dagger}\}\}=2\delta_{\alpha \beta} a_k.
\end{eqnarray}
Other parastatistical relations of this system can be derived from the above
12 basic relations by taking Hermitian conjugate or by using the generalized
Jacobi identities (8). Now we define the following 6 new operators
\begin{eqnarray}
&& \{a_k^{\dagger}, a_l\}=2M_{kl}(a), ~~~\{a_k, a_l\}=2B_{kl}(a), ~~~\{a_k,
f_{\alpha}\}=2F_{k \alpha}, \nonumber\\
&& [f_{\alpha}^{\dagger}, f_{\beta}]=2M_{\alpha \beta}(f), ~~~
[f_{\alpha}, f_{\beta}]=2B_{\alpha \beta}(f), ~~~\{a_k^{\dagger}, f_{\alpha}\}
= 2Q_{k \alpha},
\end{eqnarray}
with their Hermitian conjugate operators. Since $M_{kl}^{\dagger}(a)=
M_{lk}(a)$ and $M_{\alpha \beta}^{\dagger}(f)=M_{\beta \alpha}(f)$, actually,
there are totally 10 kinds of new operators independently, i.e.,
\par
$$
M_{kl}(a), B_{kl}(a), B_{kl}^{\dagger}(a), M_{\alpha \beta}(f),
B_{\alpha \beta}(f), B_{\alpha \beta}^{\dagger}(f), F_{k \alpha},
F_{k \alpha}^{\dagger}, Q_{k \alpha}, Q_{k \alpha}^{\dagger}.
$$
Using these new operators one can rewrite (12) as
\begin{eqnarray}
&& [a_k, M_{lm}(a)]=\delta_{kl} a_m, ~~~[a_k, M_{\alpha \beta}(f)]=0,~~~
[a_k, B_{lm}(a)]=0, \nonumber\\
&& [f_{\alpha}, M_{kl}(a)]=0, ~~~[f_{\alpha}, M_{\beta \gamma}(f)]=
\delta_{\alpha \beta} f_{\gamma}, ~~~[f_{\alpha}, B_{\beta \gamma}(f)]=0,
\nonumber\\
&& [a_k, F_{l \alpha}]=0, ~~~[a_k, Q_{l \alpha}]=\delta_{kl} f_{\alpha}, ~~~
[a_k, Q_{l \alpha}^{\dagger}]=0, \nonumber\\
&& \{f_{\alpha}, F_{k \beta}\}=0, ~~~\{f_{\alpha}, Q_{k \beta}\}=0, ~~~
\{f_{\alpha}, Q_{k \beta}^{\dagger}\}=\delta_{\alpha \beta} a_k.
\end{eqnarray}
Furthermore, by virtue of (13), (14) and (8), one can derive the following
closed algebraic relations satistied by the new operators
\begin{eqnarray}
&& [M_{kl}(a), M_{mn}(a)]=\delta_{ml} M_{kn}(a) - \delta_{kn} M_{ml}(a),
\nonumber\\
&& [M_{kl}(a), B_{mn}(a)]=-\delta_{km} B_{ln}(a) - \delta_{kn} B_{ml}(a),
\nonumber\\
&& [B_{kl}(a), B_{mn}^{\dagger}(a)]=\delta_{mk} M_{nl}(a) +
\delta_{nl} M_{mk}(a) + \delta_{ml} M_{nk}(a) + \delta_{nk} M_{ml}(a),
\nonumber\\
&& [B_{kl}(a), B_{mn}(a)]=0;   
\end{eqnarray}
\par
\begin{eqnarray}
&& [M_{\alpha \beta}(f), M_{\sigma \rho}(f)]=\delta_{\sigma \beta}
M_{\alpha \rho}(f) - \delta_{\alpha \rho} M_{\sigma \beta}(f), \nonumber\\
&& [M_{\alpha \beta}(f), B_{\sigma \rho}(f)]=-\delta_{\alpha \sigma}
B_{\beta \rho}(f) -\delta_{\alpha \rho} B_{\sigma \beta}(f), \nonumber\\
&& [B_{\alpha \beta}(f), B_{\sigma \rho}^{\dagger}(f)]=
- \delta_{\sigma \alpha} M_{\rho \beta}(f)
- \delta_{\rho \beta} M_{\sigma \alpha}(f)
+ \delta_{\sigma \beta} M_{\rho \alpha}(f)
+ \delta_{\rho \alpha} M_{\sigma \beta}(f), \nonumber\\
&& [B_{\alpha \beta}(f), B_{\sigma \rho}(f)]=0;   
\end{eqnarray}
\par
\begin{eqnarray}
&& [M_{kl}(a), M_{\alpha \beta}(f)]=0, ~~~[M_{kl}(a), B_{\alpha \beta}(f)]=0,
~~~[M_{\alpha \beta}(f), B_{kl}(a)]=0,  \nonumber\\
&& [B_{kl}(a), B_{\alpha \beta}(f)]=0, ~~~
[B_{kl}(a), B_{\alpha \beta}^{\dagger}(f)]=0;
\end{eqnarray}
\par
\begin{eqnarray}
&& [M_{kl}(a), F_{m \alpha}]= -\delta_{km} F_{l \alpha}, ~~~
[M_{\alpha \beta}(f), F_{k \gamma}]= -\delta_{\alpha \gamma} F_{k \beta},
\nonumber\\
&& [B_{kl}(a), F_{m \alpha}]=0, ~~~[B_{kl}^{\dagger}(a), F_{m \alpha}]=
- \delta_{km} Q_{l \alpha} - \delta_{lm} Q_{k \alpha}, \nonumber\\
&& [B_{\alpha \beta}(f), F_{k \gamma}]=0, ~~~[B_{\alpha \beta}^{\dagger}(f),
F_{k \gamma}]= - \delta_{\beta \gamma}Q_{k \alpha}^{\dagger} +
\delta_{\alpha \gamma} Q_{k \beta}^{\dagger}, \nonumber\\
&& [M_{kl}(a), Q_{m \alpha}]= \delta_{lm} Q_{k \alpha}, ~~~
[M_{\alpha \beta}(f), Q_{k \gamma}]= - \delta_{\alpha \gamma} Q_{k \beta},
\nonumber\\
&& [B_{kl}(a), Q_{m \alpha}]= \delta_{km} F_{l \alpha}
+ \delta_{lm} F_{k \alpha}, ~~~[B_{kl}^{\dagger}(a), Q_{m \alpha}]=0,
\nonumber\\
&& [B_{\alpha \beta}^{\dagger}(f), Q_{k \gamma}]= \delta_{\alpha \gamma}
F_{k \beta}^{\dagger} - \delta_{\beta \gamma} F_{k \alpha}^{\dagger}, ~~~
[B_{\alpha \beta}(f), Q_{k \gamma}]=0;
\end{eqnarray}
\par
\begin{eqnarray}
&& \{F_{k \alpha}, F_{l \beta}\}=0, ~~~\{Q_{k \alpha}, Q_{l \beta}\}=0, ~~~
\{F_{k \alpha}, F_{l \beta}^{\dagger}\}= \delta_{\beta \alpha} M_{lk}(a) -
\delta_{lk} M_{\beta \alpha}(f),   \nonumber\\
&& \{Q_{k \alpha}, Q_{l \beta}\}=0, ~~~\{Q_{k \alpha}, Q_{l \beta}^{\dagger}\}
= \delta_{\beta \alpha} M_{kl}(a) + \delta_{kl} M_{\beta \alpha}(f),
\nonumber\\
&& \{F_{k \alpha}, Q_{l \beta}\}= \delta_{kl} B_{\alpha \beta}(f), ~~~
\{F_{k \alpha}, Q_{l \beta}^{\dagger}\}= \delta_{\alpha \beta} B_{kl}(a),
\end{eqnarray}
with their Hermitian conjugate ones. After observing these relations carfully,
one can see that the 10 new operators form the usual $Z_2$ graded Lie algebra,
whose Bose subspace includes the operators $M_{kl}(a), B_{kl}(a),
B_{kl}^{\dagger}(a), M_{\alpha \beta}(f), B_{\alpha \beta}(f),
B_{\alpha \beta}^{\dagger}(f)$, and whose Fermi subspace includes the
operators $F_{k \alpha}, F_{k \alpha}^{\dagger}, Q_{k \alpha},
Q_{k \alpha}^{\dagger}$. Moreover, if considering the whole algebraic
relations (13-19) together, one can find that actually the following 14
operators ($a_k, a_k^{\dagger}, f_{\alpha}, f_{\alpha}^{\dagger}, M_{kl}(a),
B_{kl}(a), B_{kl}^{\dagger}(a), M_{\alpha \beta}(f), B_{\alpha \beta}(f),
B_{\alpha \beta}^{\dagger}(f), F_{k \alpha}, F_{k \alpha}^{\dagger},
Q_{k \alpha}, Q_{k \alpha}^{\dagger}$) form a $Z_{2,2}$ graded Lie algebraic
system according to the product rule (7) (we may call it para-Lie
superalgebraic system), the oprators of whose four subspaces are showing in
the Table 2. Obviously, from the point of view of the algebraic structure,
this para-Lie superalgebraic system is exactly equivalent to the parasystem
consisting of the four kinds of operators $a_k, a_k^{\dagger}, f_{\alpha},
f_{\alpha}^{\dagger}$ according to the trilinear or double commutaion
relations (12). Thus we arrive at a conclusion: The parastatistical algebraic
relations (12) are equivalent to the para-Lie superalgebraic relations (13-19)
on the algebraic structure, and the latter is the algebraic relations of
$z_{2,2}$ graded Lie algebra mathematically. This conclusion can be written as
following theorem
\par
{\bf Theorem}: {\sl The algebraic structure of parastatistics is a $Z_{2,2}$
graded Lie algebra}.

\section{Symmetries and supersymmetries of a para-Lie superalgebraic system}

>From the point of view of the para-Lie superalgebra or the $Z_{2,2}$ graded
Lie algebra, it is easily to analyse the symmetries and supersymmetries of a
paraparticle system as follows:
\par
(1) From (15) and (16) it is clear that the two subset ($M_{kl}(a), B_{kl}(a),
B_{kl}^{\dagger}(a)$) and ($M_{\alpha \beta}(f), B_{\alpha \beta}(f,)
B_{\alpha \beta}^{\dagger}(f)$) form the dynamical symmetry algebras of a pure
parabose and a pure parafermi subsystems (sp(2M,R) and so(2N,R)) respectively.
So the whole subspace $L_{bose}$ (see the Table 2) forms a Lie algebra
$sp(2M,R) \oplus so(2N,R)$.
\par
(2) From (15-19) it is clear that the whole subspace $L_{Bose} \oplus
L_{Fermi}$ form a $Z_2$ greded Lie algebra, in which $(M_{kl}(a),
M_{\alpha \beta}(f), F_{k \alpha}, F_{k \alpha}^{\dagger})$ and $(M_{kl}(a),
M_{\alpha \beta}(f), Q_{k \alpha}, Q_{k \alpha}^{\dagger})$ form two dynamical
supersymmetric algebras of the paraparticle system respectively. It should be
pointed out that for the former there is no way to construct a dynamical model
with a positive definite Hamiltonian, so it is a unaccepted supersymmetry in
physics, however, for the latter it is indeed possible to realize dynamical
sypersymmetric models with definite physical meanings.
\par
(3) From (13-19) it is clear that $(a_k, a_k^{\dagger}, M_{kl}(a), B_{kl}(a),
B_{kl}^{\dagger}(a))$ form a pure parabose statistical algebra $osp(1/2M)$, so
the whole subspace $L_{Bose} \oplus L_{Parabose}$ form a $Z_2$ graded Lie
algebra $osp(1/2M) \oplus so(2N,R)$.
\par
(4) Similarly, $(f_{\alpha}, f_{\alpha}^{\dagger}, M_{\alpha \beta}(f),
B_{\alpha \beta}(f), B_{\alpha \beta}^{\dagger}(f))$ form a pure parafermi
statistical algebra $so(2N+1,R)$, so the whole subspace $L_{Bose} \oplus
L_{Parafermi}$ form a Lie algebra $so(2N+1,R) \oplus sp(2M,R)$.
\par
Furthermore, besides the parabosons and parafermions (with $p>1$ where $p$ is
the paraststistcs order), one can include ordinary bosons and fermions ($p=1$)
in the considering system, or more detailed, put creation and annhilation
operators of the bosons (fermions) into the subspace $L_{Bose}$ ($L_{Fermi}$).
Since according to the product rule of $Z_{2,2}$, either the commutators or
anticommutators between para- and non-para- particles will be zero, the whole
algebraic structure of the enlarging system is not changed. In other words,
for the system consisting of bosons, fermions, parabosons and parafermions,
the algebraic structure of its statistical relations is still the $Z_{2,2}$
graded Lie algebra. Therefore, we can analyse the various potential
supersymmetries of such a system unifiedly within the framework of $Z_{2,2}$. 
Futher research shows that only the following three kinds of supersymmetries
(including six different cases) are possible mathematically between the four
kinds of particles:
\par
(i) The supersymmetries between boson and fermion or paraboson and
parafermion, which are realized by some fermi-like supercharges $Q_F$;
\par
(ii) The supersymmetries between boson and parafermion or paraboson and
fermion, which are realized by some parafermi-like supercharges $Q_{Pf}$;
\par
(iii) The supersymmetries between boson and paraboson or fermion and
parafermion, which are realized by some parabose-like supercharges $Q_{Pb}$.
\par
After detailed analysis we find that it is not possible to write out a
positive definite Hamiltonian in the case (iii), so we do not consider the
case (iii) further. To our knowledge, so far only part of the above six
possible supersymmetric cases has been studied in the literature. For
instance, the sypersymmetry of boson and fermion in the case (i) is studied by
ordinary supersymmetric quantum mechanics \cite{s5}, and the sypersymmetry of
boson and parafermion in the case (ii) is studied in the paraSUSY quantum
mechanics \cite{s6}. Other supersymetric cases need to be studied further.
Comparing with the ordinary statistics in which there is only one kind of
supersymmetry, i.e., boson-fermion sypersymmetry, the parastatistics allows
existance of more supersymmetries. This also indicates that the paraSUSY can
be analysed and studied more conveniently within the framework of $Z_{2,2}$
graded Lie algebra.

\section{Conclusion}
\par
The concept of $Z_{2,2}$ graded Lie algebra is introduced in this paper, which
has intrinsic connection with the parastatistics and the paraSUSY. The
$Z_{2,2}$ graded Lie algebra can unify not only paraboson and parafermion, but
also boson, fermion, paraboson and parafermion within one algebraic structure.
It is well-known that when the order $p$ of parastatistics goes to $1$, the
parastatistics will reduce to the ordinary statistics, as well as the parabose
and the parafermi subspaces will reduce to the ordinary bose and fermi
subspaces, so the $Z_{2,2}$ graded Lie algebra will reduce to the ordinary
$Z_2$ one. Therefore for the ordinary statistics only the $Z_2$ graded Lie
algebra is needed. However, when $p>1$, except original bose and fermi
subspaces, one has to introduce two extra parabose and parafermi subspaces
into the algebraic structure. In this sense we say that the $Z_{2,2}$ graded
Lie algebra is more complete in the structure and with higher symmetry than
the $Z_2$ one. Considering the discussion in section 2, we can also say that
the $Z_2$ graded Lie algebra (or orsinary statistics) is an one-dimensional
reduction of the $Z_{2,2}$ one (or parastatistics), and the $Z_{2,2}$ (or
parastatistics) is a two-dimensional generalization of the $Z_2$ (or ordinary
statistics). A high-dimensional system has higher symmetry and is more
complete constructionally than a low-dimensional system, this is a common
fact. Also in this sense, we may call the $Z_{2,2}$ graded Lie algebra (
para-Lie superalgebra) as "two-dimensional" $Z_2$ graded Lie algebra (ordinary
Lie superalgebra). Of course, searching Fock space representations of the
$Z_{2,2}$ graded Lie algebra and constructing concrete paraSUSY dynamical
models are very important and urgent problems. We will present the Fock
representation for a system with one-mode parabose and one-mode parafermi
degrees of freedom and discuss relevant paraSUSY dynamical model in a separate
paper.


\begin{thebibliography}{99}
\bibitem{s1} Green H S, {\sl A Generalized Method of Field Quantization},
{\bf Phys. Rev.} 1953 {\bf 90}: 270-274.
\bibitem{s2} Wilczek F, {\sl Fracrional Statistics and Anyon
Superconductivity}, Singapore: World Scientific, 1990.
\bibitem{s3} Beckers J and Debergh N, {\sl Parastatistics and
Supersymmetry in Quantum Mechanics}, {\bf Nucl. Phys.} 1990 {\bf B340}: 767-
776.
\bibitem{s4} Sohnius M F, {\sl Introducing  Supersymmetry}, {\bf Phys. Rep.}
1985 {\bf 128}: 58-69.
\bibitem{s5} Witten E, {\sl Dynamical Breaking of Supersymmetry}, {\bf Nucl.
Phys.} 1981 {\bf B188}: 513-554; Witten E, {\sl Constraints on Supersymmetry
Breaking}, {\bf Nucl. Phys.} 1982 {\bf B202}: 253-316.
\bibitem{s6} Rubakov V A and Spiridonov V P, {\sl Parasupersymmetric Quantum
Mechanics}, {\bf Mod. Phys. Lett.} 1988 {\bf 3}(14): 1337-1347.
\end{thebibliography}
\end{document}